\documentclass[preprint,12pt]{elsarticle}

\usepackage{graphicx}

\usepackage{amssymb}

\usepackage{amsmath}

\begin{document}

\begin{frontmatter}



\title{%
  Fourier Analysis of the Parametric Resonance in Neutrino Oscillations
}


\author[su]{Masafumi~Koike\corref{cauth}}
\cortext[cauth]{Corresponding author.}
\ead{koike@krishna.th.phy.saitama-u.ac.jp}
\author[wu]{Toshihiko~Ota}
\ead{Toshihiko.Ota@physik.uni-wuerzburg.de}
\author[su]{Masako~Saito}
\ead{msaito@krishna.th.phy.saitama-u.ac.jp}
\author[su]{Joe~Sato}
\ead{joe@phy.saitama-u.ac.jp}

\address[su]{%
  Physics Department, Saitama University,
  Sakura-ku, Saitama, Saitama 338-8570, Japan
}
\address[wu]{%
  Institut f\"{u}r theoretische Physik und Astrophysik,
  Universit\"{a}t W\"{u}rzburg, D-97074 W\"{u}rzburg, Germany
}

\begin{abstract}
  Parametric enhancement of the appearance probability of the neutrino
  oscillation under the inhomogeneous matter is studied.
  Fourier expansion of the matter density profile leads to a simple resonance
  condition and manifests that each Fourier mode modifies the energy spectrum of
  oscillation probability at around the corresponding energy; below the MSW
  resonance energy, a large-scale variation modifies the spectrum in high
  energies while a small-scale one does in low energies.
  In contrast to the simple parametric resonance, the enhancement of the
  oscillation probability is itself an slow oscillation as demonstrated by a
  numerical analysis with a single Fourier mode of the matter density.
  We derive an analytic solution to the evolution equation on the resonance
  energy, including the expression of frequency of the slow oscillation.
\end{abstract}

\begin{keyword}
Neutrino oscillation, Parametric resonance, Fourier analysis

\PACS 14.60.Pq \sep 13.15.+g \sep 14.60.Lm


\end{keyword}

\end{frontmatter}

%
Experiments over the last decade have established the oscillation among the
neutrino flavors and narrowed the allowed value of oscillation parameters.
Experiments of the next generation aiming for precision measurements are
actively discussed, and many of them involve a neutrino oscillation under a very
long baseline penetrating through the Earth.
The matter density on such a path varies spatially, reflecting the interior
structure of the Earth.
The neutrino oscillation under the varying density of matter has been widely
studied in works including Refs.~\cite%
{KOS,1986:Ermilova,ParametricResonance,Petcov}, %
and the oscillation probability is found to be significantly enhanced in some
cases.
This enhancement can be understood in a context of the parametric
resonance~\cite{1986:Ermilova,ParametricResonance}.
These studies usually model the density profile by the step function segmented
into three pieces, among which the middle piece has a larger value than the both
ends, and reproduces the qualitative feature of the numerical
results~\cite{ParametricResonance}.
In this paper, we present an alternative formulation based on a Fourier
expansion of the matter density profile, which generalizes the sinusoidal matter
profile studied in one of the earliest studies~\cite{1986:Ermilova}.
This approach organizes the effects of the matter profile on the neutrino
oscillation by their frequencies, in terms of which the condition for the
parametric resonance to take place is given.
We will analyze the neutrino oscillation in the two-flavor framework rather than
the standard three-flavor one to make the discussion simple and clear.
We briefly explain that this simplification does not wipe the effect of the
matter density profile off the oscillation probability.
\begin{figure}
\begin{center}
\includegraphics[width=120mm]{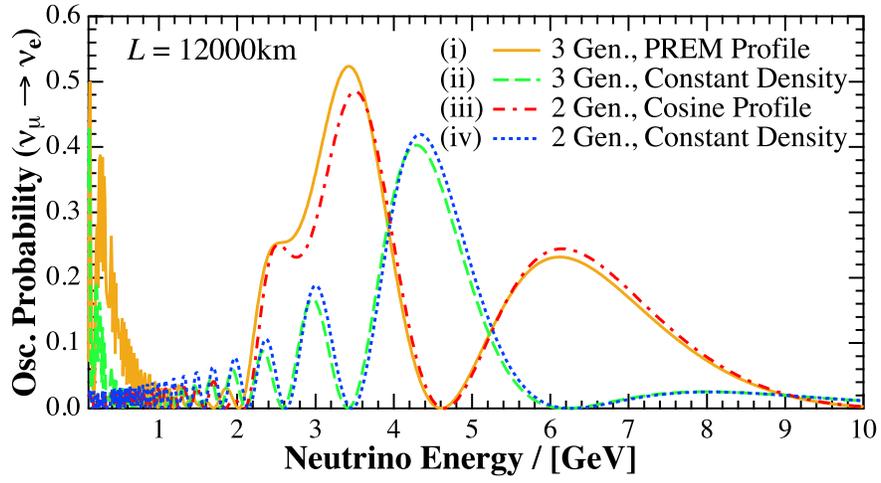}
\caption{%
  Oscillation probability $P(\nu_{\mu} \to \nu_{\textrm{e}})$ for
  $L=12\,000~\mathrm{km}$.
  The probabilities (i) and (ii) are evaluated by three-generation
  analysis and (iii) and (iv) by two-generation analysis.
  The matter density profile used %
  for (i) is the PREM profile~\cite{Dziewonski:1981xy}, %
  for (ii) and (iv) is the constant profile with $7.58\, \mathrm{g/cm^{3}} $, %
  for (iii) is the profile with 
  $[7.58 - 4.32 \cos (2\pi x/L)] \, \mathrm{g/cm^{3}}$. 
  The values of other parameters taken are described in the text. %
}
\label{fig1}
\end{center}
\end{figure}
Figure \ref{fig1} shows the $\nu_{\mu} \to \nu_{\textrm{e}}$ appearance
probability for the baseline length $L = 12\,000~\mathrm{km}$.
The curves (i) and (ii) are calculated in the three-flavor framework.
The curve (i) assumes the matter density $\rho(x)$ that varies according to the
Preliminary Reference Earth Model (PREM)~\cite{Dziewonski:1981xy}, while (ii)
assumes the constant density of $7.58 \, \mathrm{g/cm^{3}}$, which is the
average value of the PREM profile.
Other values of parameters taken here are
\begin{math}
  \delta m^{2}_{21} = 7.9 \times 10^{-5} \, \mathrm{eV}^{2}, 
  \delta m^{2}_{31} = 2.5 \times 10^{-3} \, \mathrm{eV}^{2}, 
  \sin^{2} 2\theta_{12} = 0.84, 
  \sin^{2} 2\theta_{23} = 1.0, 
  \sin^{2} 2\theta_{13} = 0.05,
\end{math}
and $\delta_{\textrm{CP}} = 0$, in the standard notations of
Ref.~\cite{Amsler:2008zz}.
We find qualitative difference between (i) and (ii), showing a major effect of
the inhomogeneity of the matter density.
The curves (iii) and (iv) are obtained from the two-flavor calculation
renormalized by $\sin^{2}\theta_{23}$.
The curve (iii) assumes the cosine density profile of the matter as $[7.58 -
4.32 \cos (2\pi x/L)] \, \mathrm{g/cm^{3}}$, where the two values are taken from
the first two Fourier coefficients of the PREM profile.
The curve (iv) assumes the constant density of $7.58 \, \mathrm{g/cm^{3}}$.
The mass-square difference and the mixing angle are taken in common as $\delta
m^{2} = 2.5 \times 10^{-3} \, \mathrm{eV}^{2}$ and $\sin^{2} 2\theta = 0.05$.
These two curves again show the significant qualitative difference from each
other.
Meanwhile, we find the curves (i) and (iii) are quite similar except in
the low energies, and so are (ii) and (iv).
This suggests the following two points:
first, the two-generation analysis is sufficient to analyze the essence of the
matter profile effect; secondly, the Fourier analysis of the matter density
profile is effective in analyzing its effect.

We study the parametric resonance of the two-flavor neutrino oscillation from
$\nu_{\mu}$ to $\nu_{\textrm{e}}$.
The evolution equation in matter is given by
\begin{subequations}
\begin{gather}
  \mathrm{i} \frac{\mathrm{d}}{\mathrm{d} x}
  \begin{pmatrix} \nu_{\textrm{e}}(x) \\ \nu_{\mu}(x) \end{pmatrix}
  =
  H(x)
  \begin{pmatrix} \nu_{\textrm{e}}(x) \\ \nu_{\mu}(x) \end{pmatrix} \, ,
  \\
  H(x)
  =
  \frac{1}{4E}
  \begin{pmatrix}
    - \delta m^{2}\cos 2\theta + 2 a(x) & \delta m^{2}\sin 2\theta \\ 
      \delta m^{2}\sin 2\theta & \delta m^{2}\cos 2\theta
  \end{pmatrix} \, ,
\end{gather}%
\label{eq:evolution_eq}%
\end{subequations}
where $E$
is the energy of neutrinos 
and
\begin{math}
  a(x)
  =
  2\sqrt{2} E G_{\textrm{F}} N_{\textrm{A}} Y_{\textrm{e}}
  \rho(x) / [1.0 \, \mathrm{g/mol}]
\end{math}
accounts for the matter effect. 
Here $G_{\textrm{F}}$, $N_{\textrm{A}}$, and $Y_{\textrm{e}}$ are the Fermi
constant, the Avogadro constant, and the proton-to-nucleon ratio, respectively.
We reduce Eq.~(\ref{eq:evolution_eq}) to a differential equation of
$\nu_{\textrm{e}}(\xi)$ in a dimensionless form as
\begin{equation}
\begin{split}
  &\nu_{\textrm{e}}''(\xi)
  + \mathrm{i} \Delta_{\textrm{m}}(\xi)
  \nu_{\textrm{e}}'(\xi)
  \\ &\hspace*{5mm} +
  \Bigl[
    \frac{\Delta^{2}}{4}
    - \frac{\Delta \cos 2\theta}{2} \Delta_{\textrm{m}}(\xi)
    + \mathrm{i} \Delta_{\textrm{m}}'(\xi)
  \Bigr] \nu_{\textrm{e}}(\xi)
  = 0 \, ,
\end{split}
\label{eq:nu-diff-eq-dimless}
\end{equation}
where $\xi \equiv x/L$ with baseline length $L$,
$\Delta_{\textrm{m}}(\xi) \equiv a(\xi)L/2E$, and
$\Delta \equiv \delta m^{2} L/2E$.
We eliminate the first-derivative term using
\begin{equation}
  z(\xi) =
  \nu_{\textrm{e}}(\xi)
  \exp \Bigl[
    \frac{1}{2} \int_{0}^{\xi} \mathrm{d}s \,
    \mathrm{i} \Delta_{\textrm{m}}(s)
  \Bigr] \, ,
\end{equation}
and then expand the matter density $\rho(\xi)$ into a Fourier series as
\begin{equation}
  \rho(\xi)
  = \rho_{0} + \sum_{n = 1}^{\infty} \rho_{n} \cos 2n \pi \xi,
\end{equation}
where $\rho_{0}$ is the average density~\cite{KOS}.
Accordingly we expand
\begin{equation}
  \Delta_{\textrm{m}}(\xi) =
  \Delta_{\textrm{m}0} + 
  \sum_{n = 1}^{\infty} \Delta_{\textrm{m} n} \cos 2n \pi \xi
\label{eq:deltaDelta_Fourier}  
\end{equation}
to obtain
\begin{subequations}
\begin{align}
  z''(\xi) +
  \biggl[
    \omega^{2}
    + \! \sum_{n = 1}^{\infty} u_{n}(\xi)
    + \! \sum_{n = 2}^{\infty} \sum_{l = 1}^{n - 1} v_{nl}(\xi)
  \biggr]
  z(\xi) = 0 \, ,
\end{align}
which we read as an equation of a harmonic oscillator, where
\begin{equation}
  \omega^{2}
  =
  \frac{1}{4}
  \Bigl[
    (\Delta_{\textrm{m}0} - \Delta \cos 2\theta)^{2}
    + \Delta^{2} \sin^{2} 2\theta
  \Bigr]
  +
  D
\end{equation}
with $D = \sum_{n = 1}^{\infty} \Delta_{\textrm{m}n}^{2}/8$
gives the natural frequency of the system, while
\begin{align}
  u_{n}(\xi)
  & =
  \alpha_{n} \cos 2n\pi\xi
  - \mathrm{i} \beta_{n} \sin 2n\pi\xi
  + \gamma_{n} \cos 4n\pi\xi,
  \\
  v_{nl}(\xi)
  & =
  \frac{\Delta_{\textrm{m}n} \Delta_{\textrm{m}l}}{4}
  \bigl[ \cos 2(n \! + \! l)\pi\xi + \cos 2(n \! - \! l)\pi\xi \bigr],
\end{align}%
\label{eq:diff-eq-z-2}%
\end{subequations}
with
\begin{subequations}
\begin{align}
  \alpha_{n}
  & =
  \Delta_{\textrm{m} n}(\Delta_{\textrm{m} 0} - \Delta \cos 2\theta)/2 \, ,
  \\
  \beta_{n} & = n \pi \Delta_{\textrm{m} n} \, ,
  \\
  \gamma_{n} & = \Delta_{\textrm{m} n}^{2}/8
\end{align}%
\end{subequations}
give oscillating potentials. 
The initial conditions $\nu_{\textrm{e}}(0) = 0$ and $\nu_{\mu}(0) = 1$ become
$z(0) = 0$ and $z'(0) = - \mathrm{i} (\Delta/2) \sin 2\theta$.
We recall that a harmonic oscillator with an oscillating potential is led to a
parametric resonance when the frequency of the oscillating potential is, in the
simplest case, twice the natural frequency of the
system~\cite{Landau:Mechanics}.
Applying this resonance condition to our system, we find that the $n$-th Fourier
mode of the matter profile, whose frequency is $2 n \pi$, gives rise to the
parametric resonance of the neutrino oscillation when $2n \pi = 2 \omega$.
This resonance condition reads $E = E_{n}^{(\pm)}$ with
\begin{equation}
  E_{n}^{(\pm)} =
  \frac{ \delta m^{2} L/2}{
    \Delta_{\textrm{m} 0} \cos 2\theta
    \pm \sqrt{
      4 n^{2} \pi^{2}
      - \Delta_{\textrm{m} 0}^{2} \sin^{2} 2\theta
      - D
    }
  } \, ,
\label{eq:parametric_resonance_energy}
\end{equation}
where $E_{n}^{(+)}$ lies below the MSW resonance while $E_{n}^{(-)}$ lies above.
The energy spectrum of the appearance probability at this energy hits the local
minimum, which we call $n$-th dip, as is evident from the initial condition
$\nu_{\textrm{e}}(0) = 0$ and the condition $\omega = n \pi$.
To summarize, %
each Fourier mode of the matter density is expected to resonate the appearance
probability of neutrinos at around the $n$-th dip of the energy spectrum.

\begin{figure}
\begin{center}
  \includegraphics[width=120mm]{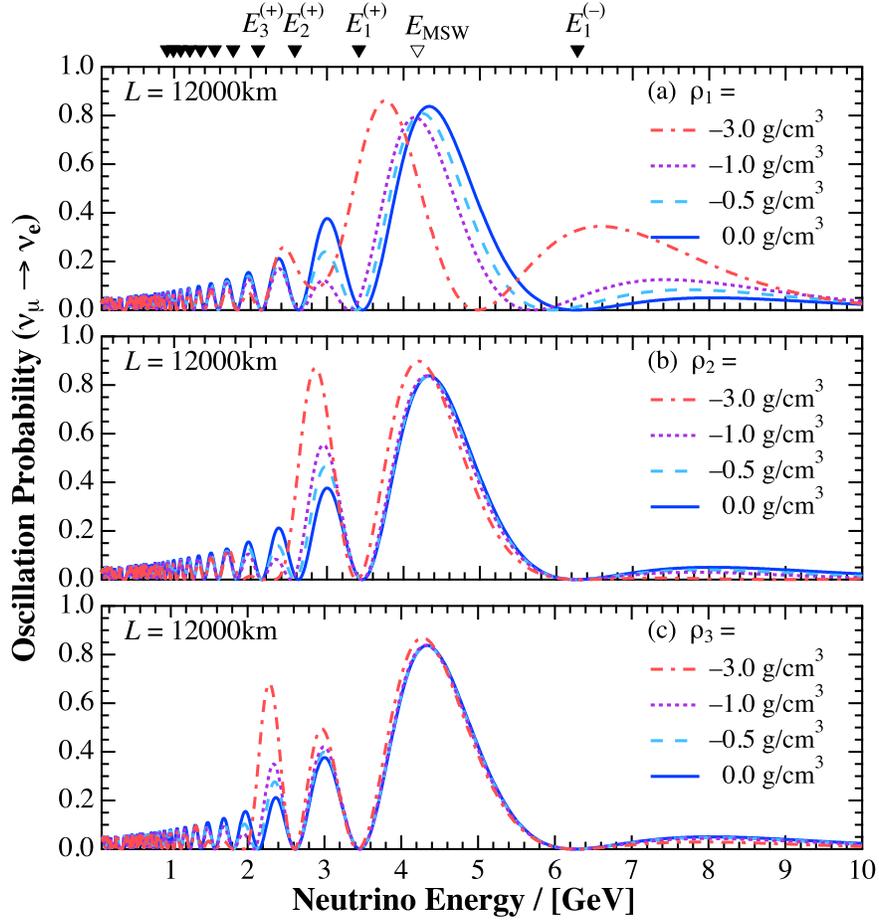}
  \caption{%
    Oscillation probability $P(\nu_{\mu} \to \nu_{\textrm{e}})$ for $\rho_{n} = 0.0
    \, \mathrm{g/cm^{3}}, -0.5 \, \mathrm{g/cm^{3}}, -1.0 \, \mathrm{g/cm^{3}},$
    and $-3.0~\mathrm{g/cm^{3}}$ with (a) $n = 1$, (b) $n = 2$ and (c) $n = 3$.
    The average density is fixed to $\rho_{0} = 7.58 \, \mathrm{g/cm^{3}}$ and
    other parameters are taken same as for (iii) and (iv) in Fig.~\ref{fig1}.%
  }
  \label{fig2}
\end{center}
\end{figure}

We elucidate the effect of the $n$-th Fourier mode by considering a simplified
model where the matter density profile consists only of the $n$-th mode around
the average density.
The matter effect in this case is described by
\begin{math}
  \Delta_{\textrm{m}}(\xi) =
  \Delta_{\textrm{m} 0} + \Delta_{\textrm{m} n} \cos 2n \pi \xi,  
\end{math}
and Eq.~(\ref{eq:diff-eq-z-2}) is reduced to
\begin{subequations}
\begin{align}
  z''(&\xi) 
      + \bigl(
        \omega_{0}^{2}
        + \alpha_{n} \cos 2n \pi \xi 
        - \mathrm{i} \beta_{n} \sin 2n \pi \xi \nonumber \\
   &\hspace*{3.2cm}  + \gamma_{n} \cos 4n \pi \xi
     \bigr) z(\xi) = 0,
   \label{eq:Hill-3term-eq} \\
   \omega_{0}^{2} 
      &= \frac{1}{4} 
          \Bigl[ ( \Delta_{\textrm{m} 0} \!-\! \Delta \cos 2\theta )^{2}
                  + \Delta^{2} \sin^{2} 2\theta
                  + \frac{\Delta_{\textrm{m} n}^{2}}{2} 
          \Bigr] \, ,
\end{align}%
\label{eq:parareso_diff-eq}%
\end{subequations}
which is free from the $v_{nl}$ terms.
This is Hill's equation, which is a generalization of Mathieu's equation that
appears in typical parametric resonance problems~\cite{HillEquation}.
To verify the discussion in the previous paragraph, we numerically solve
Eq.~(\ref{eq:parareso_diff-eq}) and show in Fig.~\ref{fig2} the oscillation
probability $P(\nu_{\mu} \to \nu_{\textrm{e}})$ for the various values of
$\rho_{1}$ in (a), $\rho_{2}$ in (b), and $\rho_{3}$ in (c).
The average density is fixed to $\rho_{0} = 7.58 \, \mathrm{g/cm^{3}}$ and 
other parameters are taken same as for (iii) and (iv) in Fig.~\ref{fig1}.
The graphs indicate that the first, the second, and the third Fourier mode of
the matter density profile indeed modify the appearance probability around
the first, the second, and the third dip, at which $\omega_{0}$ equals to $\pi$,
$2\pi$, and $3\pi$, respectively.

\begin{figure}
\begin{center}
\includegraphics[width=120mm]{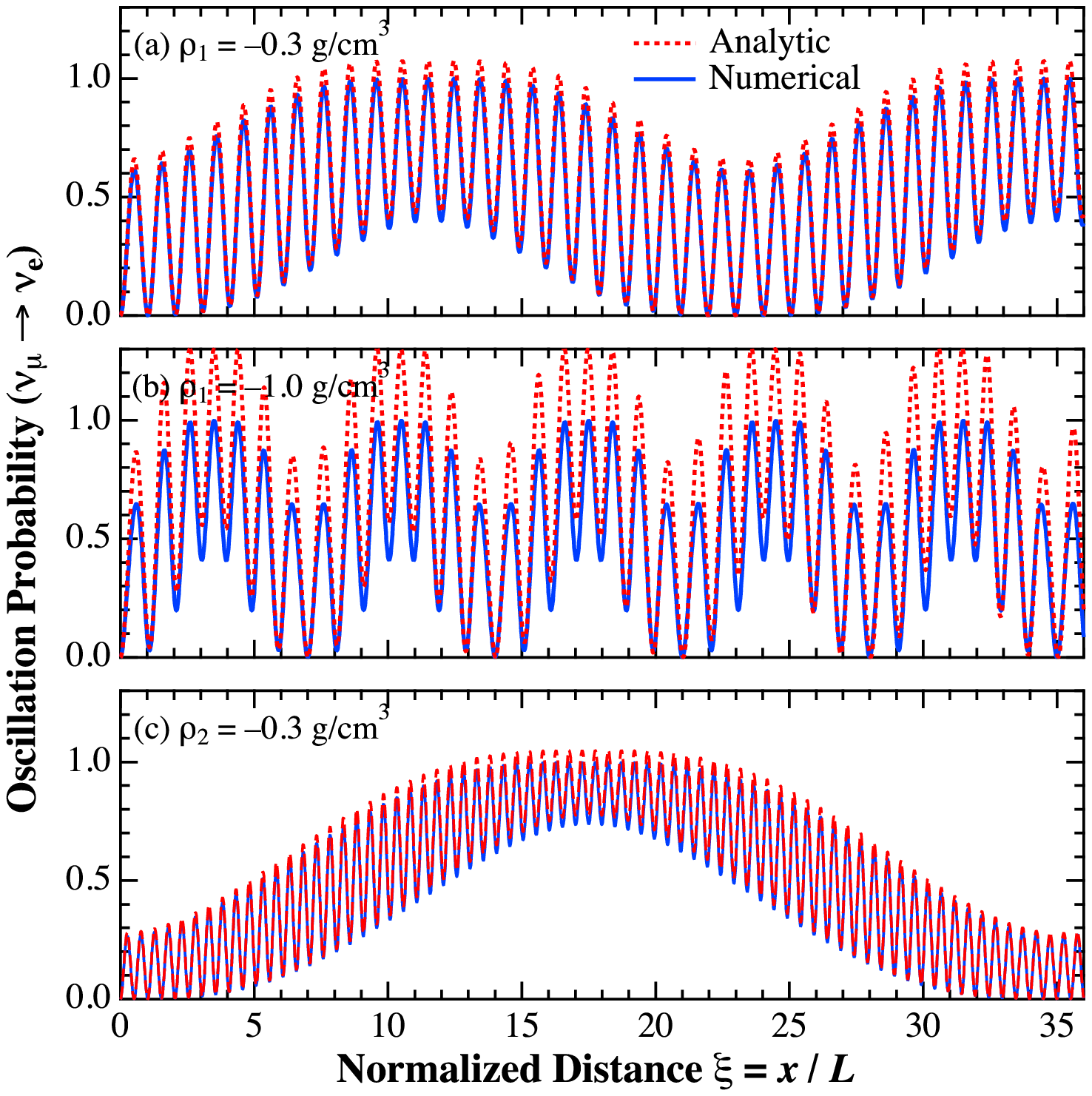}
\caption{%
  Oscillation probability $P(\nu_{\mu} \to \nu_{\textrm{e}})$ 
  beyond $\xi = 1$.
  The solid curve is given by the numerical calculation and 
  the dotted one by using the formula of Eq.~(\ref{eq:parareso_solution_z}).
  The matter profile consists of
  $\rho_{1}= - 0.3 \, \mathrm{g/cm^{3}}$ in (a), 
  $\rho_{1}= - 1.0 \, \mathrm{g/cm^{3}}$ in (b), and 
  $\rho_{2}= - 0.3 \, \mathrm{g/cm^{3}}$ in (c). 
  Neutrino energy is fixed at $E_{1}^{(+)}$ in (a) and (b) and
  $E_{2}^{(+)}$ in (c).
  The values of other parameters are same as in Fig.~\ref{fig2}.
}
\label{fig3}
\end{center}
\end{figure}

We further confirm explicitly the presence of the resonant behavior.
For simplicity, we take the energy of the neutrino at the resonance energy
$E_{n}^{(+)}$ below the MSW resonance, while the following analysis is equally
applicable to the case of $E = E_{n}^{(-)}$.
We consider the propagation of neutrinos through a repetitive density profile
and calculate $P(\nu_{\mu} \to \nu_{\textrm{e}})$ as a function of $\xi$ up to
$\xi > 1$ so that the resonant behavior becomes evident.
The result is shown in Fig.~\ref{fig3}, where
(a) $\rho_{1} = -0.3 \, \mathrm{g/cm^{3}}$, 
(b) $\rho_{1} = -1.0 \, \mathrm{g/cm^{3}}$, and 
(c) $\rho_{2} = -0.3 \, \mathrm{g/cm^{3}}$. 
The energy of neutrino is accordingly fixed at $E_{1}^{(+)}$ in (a) and (b) and
$E_{2}^{(+)}$ in (c), while other parameters are taken common and same as in
Fig.~\ref{fig2}.
The dotted curves, evaluated by the analytic formula of
Eq.~(\ref{eq:parareso_solution_z}), will be discussed later.
We find that the repetitive matter profile indeed raises the oscillation
probability as expected from the parametric enhancement.
Unlike the resonance in the classical dynamics, however, the unitarity bound
prevents the probability from the divergence.
The enhanced probability turns to decrease after reaching its peak, and thereby
lies on a slow oscillation so that the probability consists of the fast and the
slow oscillation.

We derive an analytic formula of this parametric-enhanced oscillation.
We still keep $E = E_{n}^{(+)}$ here so that $\omega_{0} = n \pi$ and $\Delta =
\delta m^{2}L / 2E_{n}^{(+)}$,
and solve Eq.~(\ref{eq:parareso_diff-eq}) as
\begin{equation}
  z(\xi)
  = A(\xi) \cos n \pi \xi + B(\xi) \sin n \pi \xi \, , 
\end{equation}
where $A(\xi)$ and $B(\xi)$ depend weakly on $\xi$. 
We then have
\begin{equation}
\begin{split}
  &
  \Bigl[ 
    A''(\xi) + 2n \pi B'(\xi) + \frac{\alpha_{n}}{2} A(\xi) 
    - \frac{\mathrm{i}\beta_{n}}{2} B(\xi)
  \Bigr]
  \cos n \pi \xi
  \\ & \hspace{-2mm}
  +
  \Bigl[ 
    B''(\xi)
    - 2n \pi A'(\xi)
    - \frac{\alpha_{n}}{2} B(\xi) 
    - \frac{\mathrm{i}\beta_{n}}{2} A(\xi)
  \Bigr] 
  \sin n \pi \xi
  \\ & \hspace{-2mm}
  + \textrm{(faster-oscillating terms)}
  = 0 \, .
\label{eq:parareso_diff-eq_expanded}  
\end{split}
\end{equation}
We neglect the faster-oscillating terms and assume
\begin{math}
  A(\xi), B(\xi) \propto \mathrm{e}^{\mathrm{i} \Omega \xi}
\end{math}
in accordance with Floquet's theorem~\cite{Akhmedov:2000js} to obtain
\begin{subequations}
\begin{align}
   \Bigl( \Omega^{2} - \frac{1}{2} \alpha_{n} \Bigr) A(\xi) 
  - \mathrm{i} \Bigl( 2n \pi \Omega - \frac{1}{2} \beta_{n} \Bigr) B(\xi) 
  = 0 \, ,
  \\
  \mathrm{i} \Bigl( 2n \pi \Omega + \frac{1}{2} \beta_{n} \Bigr) A(\xi)
  + \Bigl( \Omega^{2} + \frac{1}{2} \alpha_{n} \Bigr) B(\xi)
  =0 \, .
\end{align}%
\end{subequations}
Hence $\Omega$ is given by 
\begin{equation}
\begin{split}
  \Omega
  & =
  \pm \sqrt{2} n \pi
  \Biggl[
    1 -
    \sqrt{
      1 -
      \frac{\beta_{n}^{2} - \alpha_{n}^{2}}{(2n \pi)^{4}}
    } \,
  \Biggr]^{1/2} \, ,
\end{split}  
\label{eq:lambda}  
\end{equation}
which is real since
\begin{math}
  \omega_{0} = n\pi = \beta_{n} / \Delta_{\textrm{m} n}
\end{math}
and
\begin{math}
   \beta_{n}^{2} - \alpha_{n}^{2}
   = \Delta_{\textrm{m} n}^{2}
      ( 2\Delta^{2} \sin^{2} 2\theta + \Delta_{\textrm{m} n}^{2} ) /8
    > 0.
\end{math}
Thus $A(\xi)$ and $B(\xi)$ oscillate with the frequency $\Omega$, and give
$z(\xi)$ the slow oscillation mentioned above.
The oscillation probability can be enhanced by this slow oscillation while it
does not diverge due to its unitarity, in contrast to usual parametric
resonances.
When $\sqrt{\beta_{n}^{2} - \alpha_{n}^{2}}/(4 n \pi) \ll 1$, 
we obtain
\begin{equation}
 | \Omega |
  \simeq \frac{\sqrt{\beta_{n}^{2} - \alpha_{n}^{2}}}{4 n \pi}
  =
  \frac{\Delta_{\textrm{m} n}}{8 n \pi}
  \sqrt{ \Delta^{2} \sin^{2} 2\theta + \frac{1}{2} \Delta_{\textrm{m} n}^{2}}
  \, .
  \label{eq:Omega-Approx}
\end{equation}
Taking the lowest order in $\Delta_{\mathrm{m}n}$ of
Eq.~(\ref{eq:Omega-Approx}), we reproduce $\Omega_{n}$ in
Ref.~\cite{1986:Ermilova} with $2 \bar{\omega} = n \omega$ and $n = 1$.

The solution of Eq.~(\ref{eq:parareso_diff-eq}) 
with the initial conditions $\nu_{\mathrm{e}}(0) = 0$ and $\nu_{\mu}(0) = 1$
is expressed in terms of the oscillations with the frequencies $\omega_{0} =
n\pi$ and $\Omega$ as
\begin{equation}
\begin{split}
  &
  z(\xi)
  = 
  \frac{\mathrm{i}\Delta \sin 2\theta}
       {2\Omega ( 4 n^{2}\pi^{2} - \alpha_{n} \! - 2\Omega^{2} )}
  \Bigl[
    -  \mathrm{i} \beta_{n} \sin \Omega \xi \sin n \pi \xi
    \\ &\hspace*{-2mm}
    + \! (\alpha_{n} \! + 2\Omega^{2}) \sin \Omega \xi \cos n\pi \xi
    \! - \! 4n\pi\Omega \cos \Omega \xi \sin n \pi \xi
  \Bigr].
\label{eq:parareso_solution_z}
\end{split}
\end{equation}
We calculate $|z(\xi)|^{2}$ from Eq.~(\ref{eq:parareso_solution_z}) and plot in
Fig.~\ref{fig3} by the dotted curves.
There we see that the frequency of the slow oscillation becomes larger for
smaller $n$ and larger $|\rho_{n}|$, as we can read from
Eq.~(\ref{eq:Omega-Approx}).
The frequencies of the oscillation, both of the fast and slow ones, agree well
with the numerical results.
The amplitude of the slow oscillation agrees well with the numerical result in
(a) and (c), while the disagreement between the two grows as the frequency
$\Omega$ becomes large in (b).

The slow oscillation is extracted by averaging out the fast oscillation with the
frequency $n \pi$ as
\begin{equation}
\begin{split}
  \bigl\langle |z(\xi)|^{2} \bigr\rangle
  & =
  \frac{\Delta^{2} \sin^{2} 2\theta}
       {8\Omega^{2} ( 4 n^{2}\pi^{2} - \alpha_{n} \! - 2\Omega^{2} )^{2}}
  \times \!
  \\ & \hspace*{-11mm}
  \Bigl\{ \!
    \bigl[ ( \alpha_{n} \! + 2\Omega^{2})^{2} + \beta_{n}^{2} \bigr]
    \sin^{2} \Omega\xi
    + (4n\pi\Omega)^{2} \cos^{2} \Omega\xi
  \Bigr\}.
\end{split}
\label{eq:z-avg-approx}
\end{equation}
This equation reduces to 
$\bigl\langle |z(\xi)|^{2} \bigr\rangle = \sin^{2} \Omega\xi$
when the relations  
$\Delta_{\mathrm{m}n} \ll \Delta \sin 2\theta$ 
and
$\Omega \ll \alpha_{n}, \beta_{n} \ll 1$ are satisfied.
This result agrees with Eq.~(2) of Ref.~\cite{1986:Ermilova} 
in the lowest order in $\Delta_{\mathrm{m}n}$ 
with $2 \bar{\omega} = n \omega$ and $n = 1$.

We studied the parametric resonance in neutrino oscillation
under the inhomogeneous matter density.
We expanded the matter density profile into a Fourier series and expressed the
evolution equation of the neutrinos in a form of Hill's equation,
which leads the oscillation probability to the parametric resonance.
Noting that the resonance energy is controlled by the wavelength of the
spatial variation of the matter density,
we pointed out that the $n$-th Fourier mode of the matter density modifies the
appearance probability through the resonance at the energy around the $n$-th
dip.
We verified the correspondence between the Fourier mode and the modification of
the energy spectrum of the oscillation probability.
We further confirmed the parametric enhancement of the oscillation probability
in a calculation where the matter density profile is periodically repeated.
Unlike a simple parametric resonance, the enhancement turned out to be a slow
oscillation, which respects the unitarity.
We developed an analytic formula of the oscillation probability under the
parametric enhancement.
It well reproduced the frequencies of the slow- and the fast-oscillation of the
numerical results.
The agreement on the amplitude of the slow oscillation diminishes as the
frequency of the slow oscillation gets larger.

The Fourier analysis of the parametric resonance cleared up the correspondence
between the energy of the neutrinos and the scale of the spatial variation of
the matter density.
Below the energy of the MSW resonance, a large-scale variation modifies the
oscillation spectra in high energies while a small-scale one does in low
energies.
The effect of the density variation becomes large when neutrinos propagate
through the core of the Earth.
This effect should be kept under control in analyzing the oscillation
experiments with very long baselines, including the next-generation accelerator
experiments and the atmospheric neutrino experiments.
Conversely, it can serve as a probe for tomography of the
Earth~\cite{tomography} once the resonance effect is well understood.
Systematic Fourier analyses on the parametric resonance will give a perspective
to these experiments.

This work was supported in part by Grants-in-Aid for the Ministry of Education,
Culture, Sports, Science and Technology of Japan (Nos.~17740131, 18034001 and
19010485) (J.S.), and by the Emmy Noether program of Deutsche
Forschungsgemeinschaft under contract WI2639/2-1 (T.O.).




\end{document}